\documentclass[11pt,a4paper]{article}
\usepackage{graphicx,amsmath,amssymb,booktabs,enumerate}
\usepackage{subfigure, url,cite}
\usepackage[utf8]{inputenc}
\usepackage{comment}
\usepackage{color}

\def\a{\alpha}

\def\d{\delta}
\def\e{\epsilon}

\def\m{\mu}

\def\s{\sigma}
\def\sv{\sigma v}

\def\G{\Gamma}

\def\beq{\begin{eqnarray}}
\def\eeq{\end{eqnarray}}
\def\be{\begin{eqnarray}}
\def\ee{\end{eqnarray}}

\begin{document}

\begin{flushright}
MIT-CTP/4775
\end{flushright}
\begin{center}
{\large \bf 
Self-consistent Calculation of the Sommerfeld Enhancement
}

\vskip 1.2cm
Kfir Blum$^1$, Ryosuke Sato$^{1,2}$, Tracy R.~Slatyer$^3$
\vskip 0.4cm

$^1${\it
Department of Particle Physics and Astrophysics,\\
Weizmann Institute of Science, Rehovot 7610001, Israel\\
}
$^2${\it
Institute of Particle and Nuclear Studies,\\
High Energy Accelerator Research Organization (KEK)\\
Tsukuba 305-0801, Japan\\
}
$^3${\it
Center for Theoretical Physics, Massachusetts Institute of Technology,\\
Cambridge, MA 02139, USA\\}

\vskip 1.5cm

\abstract{
A calculation of the Sommerfeld enhancement is presented and applied to the problem of s-wave non-relativistic dark matter annihilation.
The difference from previous computations in the literature is that the effect of the underlying short-range scattering process is consistently included together with the long-range force in the effective QM Schr\"odinger problem. 
Our procedure satisfies partial-wave unitarity where previous calculations fail. We provide analytic results for  some potentials of phenomenological relevance.
}
\end{center}

\section{Introduction}
The annihilation cross section of thermal relic weakly-interacting massive particle (WIMP) dark matter controls both the relic abundance through freeze-out in the early Universe as well as the prospects for indirect detection. It is important to establish to what extent the annihilation cross section at freeze-out, when WIMPs are mildly non-relativistic with thermal velocities of order $v\sim0.2$, is related to the annihilation cross section in systems where the WIMP is highly non-relativistic, including the Galaxy ($v\sim5\times10^{-4}$) or the epoch of recombination ($v\sim10^{-10}$ for a WIMP with $\sim$~TeV mass and kinetic decoupling temperature of a few MeV).

A well-known effect that can modify the annihilation cross section in the deep non-relativistic regime is the Sommerfeld effect (SE), arising when the WIMP couples to a light force mediator~\cite{sommerfeld, Hisano:2002fk, Hisano:2003ec, Hisano:2004ds}.
The SE was estimated in the literature using a factorized formula (see, \textit{e.g.} Ref.~\cite{Cirelli:2007xd, ArkaniHamed:2008qn}),
\begin{align}
\sv = \sv_0 \times S(v). \label{eq:usual_formula}
\end{align}
Eq.~(\ref{eq:usual_formula}) is computed in two steps:
\begin{enumerate}
\item $\sv_0$ is the velocity-weighted {\it hard} annihilation cross section,
computed using the short-range physics alone, ignoring the long-range potential
\item The SE factor $S(v) \equiv |\psi_0(0)|^2$ is derived from $\psi_0(x)$, the wave function for the non-relativistic two-body dark matter system,
 determined as a solution to the quantum mechanics (QM) scattering problem using a Schr\"odinger equation in which the short-range physics is omitted, with asymptotic behaviour $\psi_0(r\to\infty)\to e^{ikz}+f\frac{e^{ikr}}{r}$.
\end{enumerate} 

Eq.~(\ref{eq:usual_formula}) was used in many examples in the literature, but in certain situations it breaks down. 
A noted example is the occurrence of a bound state with zero binding energy in the long-range Schr\"odinger equation. (See, \textit{e.g.}, \S 133 in Ref.~\cite{Landau:1991wop}.)
If a bound state solution exists with binding energy $|E_b|$ much smaller than the typical potential energy $V_0$, then
the SE factor approaches the form\footnote{
Ref.~\cite{MarchRussell:2008tu} discussed this limit using a modified Breit-Wigner formula  introduced by Bethe and Placzek \cite{BethePlaczek}. See also \S 145 in Ref.~\cite{Landau:1991wop}.}
$S(v) \sim \frac{V_0}{ \mu v^2/2 + |E_b| }.$ Since $\sv_0$ approaches a constant for $s$-wave hard-annihilation process (the Bethe $1/v$ law), Eq.~(\ref{eq:usual_formula}) gives velocity-weighted cross section $\sv\propto 1/v^2$ at small $v$ for $|E_b|\to0$.
This violates parametrically the partial-wave unitarity limit~\cite{Landau:1991wop, Griest:1989wd}, $
\s \leq \pi/k^2=\pi/(\mu v)^2.$ 
A more obvious example in which Eq.~(\ref{eq:usual_formula}) fails, is when $\sv_0$ is itself not very small, such that a finite $S(v)$ can lead to $\sv$ violating the unitarity bound at finite velocity without any special resonance of parametric origin.

In the current paper we identify the source of trouble in Eq.~(\ref{eq:usual_formula}) and fix it in a simple way. 
The basic result is easy to explain: probability conservation requires that the naive SE factor $S(v)=|\psi_0(0)|^2$, given by the modulus of the wave function at the origin neglecting the short-range potential, should be replaced by $|\psi(0)|^2$, the modulus of the full wave function including the short-range interaction\footnote{
A similar situation was discussed in the context of nuclear physics,
e.g., non-relativistic proton-proton fusion \cite{Kong:1998sx, Kong:1999tw}. There, the emphasis was on resummation of strong short-range elastic scattering. Here we address the problem from a somewhat different perspective, and will mainly focus on the resummation of short-range inelastic scattering (annihilation). 
}.
Our main task in the paper will be to compute $\psi(0)$ and to show how using it instead of $\psi_0(0)$ in Eq.~(\ref{eq:usual_formula}) restores unitarity. 

Our formula for the full annihilation cross section, given in Eq.~(\ref{eq:sigma_ann1}), is only slightly more complicated than Eq.~(\ref{eq:usual_formula}). To compute the full SE we require one more quantity [denoted $T(v)$ in Eq.~(\ref{eq:sigma_ann1})] based on the long-range interaction, in addition to the usual $S(v)$; and one more quantity [the short-range elastic scattering cross section, denoted $\sigma_{sc,0}$ in Eq.~(\ref{eq:sigma_ann1})] based on the short-range interaction, in addition to the hard annihilation cross section $\sv_0$. We provide analytic expressions of the new $T(v)$ term for specific long-range potentials of interest in typical applications, including the Coulomb and well potentials and the Hulth\'en potential (approximating the Yukawa potential), for which the usual $S(v)$ term was derived elsewhere in the literature.


In Sec.~\ref{sec:formulae} we derive our main result and present a general formula for the full SE. In Sec.~\ref{ssec:vc} we show that close to a resonant point, for perturbative underlying short-range cross section, the full SE can be obtained by simple velocity shift $S(v)\to S(v+v_c)$ in Eq.~(\ref{eq:usual_formula}). The shift velocity $v_c$ contains a product of short- and long-range information that we compute explicitly for the Hulth\'en potential.    
In Sec.~\ref{sec:examples} we provide analytic results for specific potentials. We focus in particular on the Hulth\'en potential, approximating the Yukawa interaction, where we demonstrate how our SE computation conserves partial-wave unitarity. In Sec.~\ref{sec:conc} we conclude.

\section{Formulation}\label{sec:formulae}
%
The non-relativistic effective action for the dark matter two-body system is obtained
by integrating out degrees of freedom with large momentum \cite{Hisano:2002fk, Hisano:2003ec, Hisano:2004ds}.
The resulting Schr\"odinger equation gives an effective QM description for momentum $p\ll \m$,
\begin{align}
\left[-\frac{1}{2\m}\nabla^2 + V(|\mathbf{x}|) + u \d^{(3)}(\mathbf{x}) - \frac{p^2}{2\m} \right] \psi(\mathbf{x}) = 0. \label{eq:deq}
\end{align} 
Here $\mu=M/2$ is the reduced mass. 
Long-range light mediator exchange gives rise to the real, rotationally-symmetric potential $V(r)$, $r=|\mathbf x|$.
The short-range process is encoded in the term $u\d^3(\mathbf x)$ with complex parameter $u$.
We note that Eq.~(\ref{eq:deq}) is adequate for s-wave  short-range scattering, which is what we address here and is enough to establish the formalism. Generalization to higher-$l$ initial state is left for subsequent work.

Our objective is to solve Eq.~(\ref{eq:deq});
the annihilation cross section can then be read from the solution as follows. The probability current associated with $\psi(\mathbf{x})$ is $j(\mathbf{x})=\frac{1}{\m}{\rm Im}\psi^*(\mathbf{x})\nabla\psi(\mathbf{x})$. Using the Hamiltonian $H=-\frac{1}{2\m}\nabla^2 + V(|\mathbf{x}|) + u \d^{(3)}(\mathbf{x})$, and imposing steady state $\partial_t|\psi(\mathbf{x})|^2=0$, we have 
\be
\nabla\cdot j(\mathbf{x})=2{\rm Im} u|\psi(\mathbf{0})|^2\delta^{(3)}(\mathbf{x}).\ee
The full velocity-weighted annihilation cross section is given by integrating the divergence of the probability current,
\be\label{eq:full}\sv&=&-\int d^3\mathbf{x}\nabla\cdot j(\mathbf{x})=(-2{\rm Im} u)|\psi(\mathbf{0})|^2.\ee
The boundary condition on $\psi(\mathbf{x})$ is that at large $r$ it should be given by an incident plane wave plus outgoing spherical wave, $\psi(\mathbf{x})\to e^{ipz}+fe^{ipr}/r$. The factor $v=p/\mu$ on the LHS of Eq.~(\ref{eq:full}) is the relative velocity between the incoming particles at $z\to\infty$, namely, the flux of the incident wave. 
We are already in place to compare with earlier literature.
\begin{itemize}
\item Neglecting both the long- and the short-range potentials in Eq.~(\ref{eq:deq}) gives a free wave $\psi_{free}(\mathbf{x})=e^{ikz}$, leading to the usual identification ${\rm Im}u=-\frac{\sv_0}{2}$. 
\item Neglecting the short-range potential but keeping the long-range potential in Eq.~(\ref{eq:deq}) gives the solution $\psi_0(\mathbf{x})$, leading to Eq.~(\ref{eq:usual_formula}) with $S(v)=|\psi_0(\mathbf{0})|^2$.
\item The full solution $\psi(\mathbf{x})$, where we solve Eq.~(\ref{eq:deq}) retaining both the short- and the long-range potentials, gives the full SE as $|\psi(\mathbf{0})|^2$. The shortcoming of Eq.~(\ref{eq:usual_formula}) is that it fails to conserve probability by using $\psi_0$ instead of $\psi$.
\end{itemize}

We solve Eq.~(\ref{eq:deq}) following Jackiw~\cite{Jackiw:1991je} that analyzed the scattering amplitude for the QM delta-function potential.
We define the Green's function $G(\mathbf{x})$ and
the wave function $\psi_{0}(\mathbf{x})$, chosen to satisfy
\begin{align}
\left[-\frac{1}{2\m}\nabla^2 + V(|\mathbf{x}|) - \frac{p^2}{2\m} \right] G (\mathbf{x}) & =
     \frac{1}{2\m}\d^{(3)}(\mathbf{x}), \label{eq:deq_G}\\
\left[-\frac{1}{2\m}\nabla^2 + V(|\mathbf{x}|) - \frac{p^2}{2\m} \right] \psi_{0}(\mathbf{x}) &=
     0.
\end{align}
The function $\psi_{0}(\mathbf{x})$ is the regular solution to the Schr\"odinger equation that corresponds to a unit-normalized initial plane wave; that is, $\psi_{0}(\mathbf{0})$ is finite, and $\psi_0(\mathbf{x})\to e^{ipz}+f_0e^{ipr}/r$ at large $r$, where the second term represents a contribution to the scattered wave. The function
$G(\mathbf{x})$ is chosen to give an out-going spherical wave at $r\to\infty$.
Namely, the boundary condition for $G$ at infinity is written as
\begin{align}
\lim_{|\mathbf{x}|\to\infty} \left(\frac{e^{+ip|\mathbf{x}|}}{4\pi|\mathbf x|}\right)^{-1}  G (\mathbf{x}) = d_p, \label{eq:bec_G}
\end{align}
where $d_p$ is a function of $p$.

The solution of Eq.~(\ref{eq:deq}) is  
\begin{align}
\psi(\mathbf{x}) &= \psi_{0}(\mathbf{x}) - 2\m u \psi(\mathbf 0)G(\mathbf x), 
\end{align}
so consistency at the origin implies 
\be\label{eq:concic}\psi(\mathbf 0)&=&\frac{\psi_{0}(\mathbf 0)}{1 + 2\m u G(\mathbf 0)},
\ee
leading to
\begin{align}
\psi(\mathbf{x}) &= \psi_{0}(\mathbf{x})-G(\mathbf x) \psi_{0}(\mathbf 0)
\left( \frac{1}{2\m u} + G(\mathbf 0) \right)^{-1}. \label{eq:bare_solution}
\end{align}
Using Eq.~(\ref{eq:concic}) in Eq.~(\ref{eq:full}) gives us the full SE directly: ${\rm SE}=S(v)/\left|1 + 2\m u G(\mathbf 0)\right|^2$. The denominator corrects Eq.~(\ref{eq:usual_formula}) and shows in what way factorization needs to be modified. However, we are not quite there yet, because $G(\mathbf 0)$ is divergent.
This divergence is regulated with a redefinition of the parameter $u$ absorbing an infinite constant, as we now discuss.

Rewrite $G(\mathbf{x})$ as
\begin{align}
G(\mathbf{x}) = \frac{g_p(r)}{4\pi r}.
\end{align}
From Eqs.~(\ref{eq:deq_G}) and~(\ref{eq:bec_G}), $g_p(r)$ can be shown to satisfy\footnote{
To derive Eq.~(\ref{eq:bcd_for_greenfunction}), we used Eq.~(\ref{eq:deq_for_greenfunction}), $\lim_{r\to 0} r g_p'(r) = 0$ and 
\begin{align}
-\e^2 \left(\frac{d}{dr} \frac{g_p(r)}{r}\right)\biggr|_{r=\e} + \int_0^\e dr' r'\left( 2\m V(r') -p^2 \right) g_p(r') = 1. \nonumber
\end{align}
}
\begin{align}
\left[ -\frac{d^2}{dr^2} + 2\m V(r) - p^2 \right] g_p(r) &= 0, \label{eq:deq_for_greenfunction}\\
g_p(0) &= 1 \label{eq:bcd_for_greenfunction}, \\
\lim_{r\to\infty} g_p(r)  &= d_pe^{ipr}.\label{eq:bcd2_for_greenfunction}
\end{align}
We are interested in $G(\mathbf x)$ near the origin.
We limit the discussion to a long-range potential $V(r)$ that does not diverge at the origin faster than $1/r$, and
 can be expanded as 
\begin{align}\label{eq:Vexpand}
2\m V(r) = \frac{\mathcal{V}_{-1}}{r} + \mathcal{V}_0 + \mathcal{V}_1 r + \cdots.
\end{align}
This type of potential includes the Yukawa, Coulomb, and well potentials. 
Near the origin $g_p(r)$ and $G(r)$ are found as
\begin{align}
g_p(r) &= (1 + g_1 r + g_2 r^2 + \cdots) + \mathcal{V}_{-1}(r + h_2 r^2 + \cdots ) \log r, \\
G(r) &= \frac{1}{4\pi r} + \frac{\mathcal{V}_{-1}}{4\pi} \log r + \frac{g_1}{4\pi} + {\cal O}(r\log r).
\end{align}
%
%
Note that the divergent part of $G(0)$ is independent of momentum $p$. In addition, the divergence is restricted to ${\rm Re}G(0)$, while ${\rm Im}G(0)$ is finite for real $V(r)$. 

To subtract the divergence we define a renormalized coupling at reference momentum $p_0$,
\begin{align} \label{eq:kpo}
\frac{k_{p_0}}{4\pi} \equiv \frac{1}{2\m u} + {\rm Re}G_{p_0}(\mathbf 0).
\end{align}
The subscript on $G_{p_0}$ is there to emphasize that this $G$ is to be evaluated at $p_0$. 
In QFT language,
$(2\m u)^{-1}$ is a bare coupling and ${\rm Re}G_{p_0}(\mathbf 0)$ is a counter term. Deferring some subtleties to the next subsection, it is convenient to take large reference momentum $p_0$ and to define the short-range cross section $\sv_0$ as the cross section measured at $p_0$. 
%
We are left with the task of computing $k_{p_0}$ from this matching procedure; this we do in the next subsection by appealing to some more basic QM.

Before moving on, note that above we used a regularization scheme in which $G(0)$ is replaced by $G(\e)$, finally taking $\e\to 0$. Jackiw~\cite{Jackiw:1991je} took a different approach, examining both momentum cut-off and dimensional regularization. But then, Jackiw~\cite{Jackiw:1991je} was interested in the scattering amplitude for the delta function potential alone, while we are interested in the interplay between short- and long-range physics. To this end we find our procedure convenient.

\subsection{Cross section matching}
Rewrite Eq.~(\ref{eq:bare_solution}) as
\begin{align} 
\psi(\mathbf{x}) &= \psi_{0}(\mathbf{x}) - \frac{ \psi_{0}(\mathbf 0) }{k_{p_0} - {\rm Re}g_{p_0}'(0) + g_p'(0)}
\frac{  g_p(|\mathbf{x}|)}{|\mathbf{x}|}.
\label{eq:sol_deq}
\end{align}
We take the partial wave expansion as (See, \textit{e.g.}, \S 123 in Ref.~\cite{Landau:1991wop})
\begin{align}
\psi(\mathbf x) &= \psi_s(|\mathbf x|) + \psi_p(|\mathbf x|) P_1(\cos\theta) + \cdots, \\
\psi_0(\mathbf x) &= \psi_{s,0}(|\mathbf x|) + \psi_{p,0}(|\mathbf x|) P_1(\cos\theta) + \cdots,
\end{align}
where the $P_\ell$ are Legendre polynomials and $\cos\theta = z/r$.	
Considering $s$-wave scattering we define the corresponding radial wave functions
\begin{align}
\psi_s(r) = \frac{\chi_s(r)}{r},\qquad
\psi_{s,0}(r) = \frac{\chi_{s,0}(r)}{r}.
\end{align}
Since $\psi_0(0)$ is finite, the boundary condition for $\chi_{s,0}$ is $\chi_{s,0}(0) = 0$. This gives
\begin{align}
\chi_s(r) = \chi_{s,0}(r) - \frac{g_p(r) \chi'_{s,0}(0)}{k_{p_0} - {\rm Re}g_{p_0}'(0) + g_p'(0)}. \label{eq:sol_deq_chi}
\end{align}
The function $\chi_{s,0}(r)$ solves the same 2nd order differential equation as the function ${\rm Im}g_p(r)$, with the same (null) boundary condition at the origin. The other boundary condition at $r\to\infty$ only fixes the normalization. Therefore $\chi_{s,0}(r) = c{\rm Im}g_p(r)$, where $c$ is an (in general complex) constant multiplicative factor.
Thus, for large $r$,
\begin{align}\label{eq:chiS}
\chi_s(r)
&\to
\left(
\frac{  k_{p_0}  - {\rm Re}g'_{p_0}(0) + {g'_p}^*(0) }{ k_{p_0} - {\rm Re}g'_{p_0}(0) + g'_p(0) }
 \right) \frac{c d_p e^{ipr} }{2i}
- \frac{c d_p^* e^{-ipr}}{2i}.
\end{align}

Using Eq.~(\ref{eq:chiS}), the s-wave scattering amplitude is 
\begin{align}
S_0
&= \frac{d_p}{d_p^*} \frac{  k_{p_0}  - {\rm Re}g'_{p_0}(0) + {g'_p}^*(0) }{ k_{p_0} - {\rm Re}g'_{p_0}(0) + g'_p(0) }.
\end{align}	
%
The elastic scattering, annihilation, and total cross sections are (see, e.g., \S 142 in
Ref.~\cite{Landau:1991wop})
\begin{align}
\s_{sc} = \frac{4\pi}{p^2}\left| \frac{S_0-1}{2i} \right|^2,\qquad
\s_{ann} = \frac{4\pi}{p^2} \frac{1-|S_0|^2}{4},\qquad
\s_{tot} = \frac{4\pi}{p^2}{\rm Im}\frac{S_0-1}{2i}.
\end{align}
%
In terms of $g_p'(0)$ and $k_{p_0}$, $\s_{ann}$ is written as
\begin{align}
\s_{ann} &= \frac{4\pi}{p^2} \frac{ {\rm Im}k_{p_0} {\rm Im}g'_p(0) }{\left| k_{p_0} - {\rm Re}g'_{p_0}(0) + g'_p(0) \right|^2}. \label{eq:sigma_ann}
\end{align}
The elastic scattering cross section is
\begin{align}
\s_{sc} &= \frac{\pi}{p^2} \left|
\frac{d_p}{d_p^*} \frac{  k_{p_0}  - {\rm Re}g'_{p_0}(0) + {g'_p}^*(0) }{ k_{p_0} - {\rm Re}g'_{p_0}(0) + g'_p(0) } - 1
 \right|^2.
\label{eq:sigma_sc}
\end{align}
%

To isolate the short-range physics we need to consider the limit of large kinetic energy $p^2/2\m$ compared to the typical potential energy. To be precise, we define $V(r)=\mathcal{V}_{-1}/(2\m r)+\tilde{V}(r)$, where $\tilde{V}(r)$ is assumed to be regular everywhere. The large momentum limit we are interested in satisfies $p^2\gg\left|2\m\tilde{V}(r)\right|$ for all $r$. 
In this limit the solution for $g_p(r)$ that satisfies $g_p(0)=1$ at the origin, and that matches to an outgoing wave in the leading WKB approximation at large $r$, is found as
\begin{align}\label{eq:Whitt}
g_p(r) = \frac{i \mathcal{V}_{-1}}{2 p} \G\left( \frac{i \mathcal{V}_{-1}}{2 p}\right) W\left( -\frac{i \mathcal{V}_{-1}}{2 p}, \frac{1}{2}, -2ipr \right)+...
\end{align}
where $W(k,m,z)$ is the Whittaker $W$ function. The dots (...) stand for sub-leading corrections to the WKB approximation.
 For $p, p_0 \gg \mathcal{V}_{-1}$,
\begin{align}
g_p'(0) - {\rm Re}g_{p_0}'(0)
\simeq
ip + \mathcal{V}_{-1}\log\frac{p}{p_0},\;\;\;\;\;\;\;d_p\simeq1.
\end{align}
Thus
\begin{align}
\label{eq:sannVreg}\s_{ann} &\simeq \frac{4\pi}{p} \frac{ {\rm Im}k_{p_0}  }{\left|k_{p_0} + ip + \mathcal{V}_{-1}\log \frac{p}{p_0}\right|^2}, \\
\label{eq:sscVreg}\s_{sc} &\simeq \frac{ 4\pi  }{\left|k_{p_0} + ip + \mathcal{V}_{-1}\log \frac{p}{p_0}\right|^2} .
\end{align}
%

Recall that we have defined the renormalized inverse-scattering length parameter $k_{p_0}$ in Eq.~(\ref{eq:kpo}) by appealing to the Green's function at reference momentum $p_0$. We see that if we set $p=p_0$ and consider large $p_0$ w.r.t. the long-range potential, in the sense discussed above, we can use Eqs.~(\ref{eq:sannVreg}-\ref{eq:sscVreg}) to fix $k_{p_0}$, up to the sign of ${\rm Re} k_{p_0}$, from the cross section defined at $p_0$. 

In perturbative examples life is easier because\footnote{To see this, note that the parameter $k_{p_0}$ in the effective QM picture is obtained from the QFT by matching to the relativistic cross section calculation. The scaling is $k_{p_0}\sim1/(\mu u)\sim\mu/\alpha$, where $u\sim\sigma v\sim\alpha/\mu^2$ and $\alpha$ is the appropriate combination of QFT coupling constants entering the scattering cross section. In a perturbative theory $|\alpha|\ll1$ and so $|k_{p_0}|\gg\mu$. Concerning the potential term $\mathcal{V}_{-1}$, for e.g. Yukawa theory $V(r)=-\alpha e^{-mr}/r$ we have $|\mathcal{V}_{-1}|=2|\alpha|\mu\ll\mu$.} ${\rm Im}k_{p_0}$ is larger than $\m$ (and so larger than $p_0$) and $\mathcal{V}_{-1}$ is smaller than $\m$. 
Assuming ${\rm Im}k_{p_0} \gg p_0$, 
\begin{align}
\sv_0 \simeq -\frac{4\pi}{\mu} {\rm Im}k_{p_0}^{-1}\;\;\;\;\Rightarrow\;\;\;\;\m\,{\rm Im} k_{p_0}^{-1}\simeq-\frac{\m^2\sv_0}{4\pi}\label{eq:nonrel1}
\end{align}
and
\begin{align}
\s_{sc,0} \simeq 4\pi|k_{p_0}^{-1}|^2\;\;\;\;\Rightarrow\;\;\;\;\m^2\left({\rm Re} k_{p_0}^{-1}\right)^2\simeq\frac{\m^2\sigma_{sc,0}}{4\pi}-\left(\frac{\mu^2\sv_0}{4\pi}\right)^2. \label{eq:nonrel2}
\end{align}
The sign of ${\rm Re} k_{p_0}^{-1}$ is model-dependent, in the sense that positive (negative) sign corresponds to a repulsive (attractive) short-range force. 

In what follows we focus, for simplicity, on the perturbative regime where Eqs.~(\ref{eq:nonrel1}-\ref{eq:nonrel2}) are valid and our results take a particularly simple form.

We still need to compute the long-range effect encoded in $g'_p(0)$ and $d_p$. We first note that
\be\label{eq:aux1}
S(v)&=&|d_p|^2 =\frac{1}{p}{\rm Im}g_p'(0) = \left|g_p(\infty)\right|^2,
\ee
where $S(v)$ is the usual naive SE factor from Eq.~(\ref{eq:usual_formula}).
To prove this, define the Wronskian 
\begin{align}
W_p(r) \equiv g_p'(r) g_p^*(r) -  g_p'^*(r) g_p(r) ,
\end{align}
satisfying $W_p'(r) = 0$.
At the origin and at $r\to\infty$ we have 
\begin{align}
\frac{1}{2i}W_p(0) = {\rm Im}g_p'(0),\qquad
\lim_{r\to\infty} \frac{1}{2i}W_p(r) = p|d_p|^2.
\end{align}
Together with $g_p(0)=1$, derived before, this gives Eq.~(\ref{eq:aux1}).

It is useful at this point to refer again to the usual formula Eq.~(\ref{eq:usual_formula}). This should arise as the limit of our Eq.~(\ref{eq:sigma_ann}) for $|k_{p_0}| \gg |g_p'(0) - g_{p_0}'(0)|$.
Using Eq.~(\ref{eq:aux1}) we have, in this limit,
\begin{align}
\sv \simeq \left(\frac{{\rm Im}g'_p}{p} \right)\times\left(-\frac{4\pi}{\m} {\rm Im}k_{p_0}^{-1}\right) 
= \sv_0 \times S(v), 
\end{align}
as needed. 

Last we define 
\be T(v)\equiv \frac{{\rm Re}g'_p(0)-{\rm Re}g'_{p_0}(0)}{p}.\ee
In general, for arbitrary small momentum $p$, ${\rm Re}g_p'$ at the origin must be determined by solving Eq.~(\ref{eq:deq_for_greenfunction}) using the full long-range potential; this situation is similar to that of the usual $S(v)$ factor. Below we will give analytic solutions for specific examples: (i) the Hulth\'en potential, approximating the Yukawa potential; (ii) the Coulomb potential arising from a massless mediator; (iii) the well potential.

Putting everything together, we can write Eq.~(\ref{eq:sigma_ann}) as
\be
\label{eq:sigma_ann1}\sv &\simeq& \frac{\sv_0 \,S(v) }{\left| 1 +\left(\eta\sqrt{\frac{\m^2\sigma_{sc,0}}{4\pi}-\left(\frac{\mu^2\sv_0}{4\pi}\right)^2}-i\frac{\m^2\sv_0}{4\pi}\right)\left(T(v)+iS(v)\right)v \right|^2},
\ee
where $\eta=\pm1$ is the sign of ${\rm Re}k_{p_0}^{-1}$. An analogous expression can be written for the elastic scattering cross section, but here and in what follows we focus on annihilation. 

This concludes our basic calculation: the SE factor for annihilation, including regularization by the short-range process, is given as the coefficient of $\sv_0$ in Eq.~(\ref{eq:sigma_ann1}). In the next section we provide analytic computations of the SE for specific examples. Analytic calculations and approximations for the $S(v)$ term exist in several references; we add to those a calculation of the $T(v)$ term. The short-range information: $\sv_0,$ $\sigma_{sc,0}$, and the sign $\eta$ of the short-range force, should be provided according to the underlying model. For example, a thermal-freezout WIMP model would have $\sv_0\approx3\times10^{-26}$~cm$^3$/sec, etc.

\subsection{Regularising velocity near a zero energy bound state}\label{ssec:vc}
Particularly strong SE arise when the Schr\"odinger equation including the long-range potential admits a bound state solution with zero binding energy $|E_b|\to0$~\cite{Hisano:2002fk, Hisano:2003ec, Hisano:2004ds,Feng:2010zp}. At small velocity near $|E_b|\to0$, the naive SE factor scales as
\be\label{eq:Sres} S(v)\approx\frac{V_0}{(\m/2)v^2+|E_b|}\to\frac{2V_0}{\m v^2},\ee
where $V_0$ is the characteristic scale of the potential energy.  Ref.~\cite{Hisano:2002fk, Hisano:2003ec, Hisano:2004ds} noted that the singular $1/v^2$ scaling is cut-off when the short-range contact term is accounted for perturbatively in solving the Schr\"odinger equation, and that the SE saturates at some velocity $v_c$. Using Eq.~(\ref{eq:sigma_ann1}) we can verify these observations directly.

For simplicity and for the sake of comparison with Refs.~\cite{Hisano:2002fk, Hisano:2003ec, Hisano:2004ds} we neglect the real part of the inverse-scattering length induced by the short-range interaction. This amounts to setting $\sigma_{sc,0}=\left(\frac{\mu^2\sv_0}{4\pi}\right)\sv_0$ in Eq.~(\ref{eq:sigma_ann1}).
Plugging the asymptotic form Eq.~(\ref{eq:Sres}) into Eq.~(\ref{eq:sigma_ann1}) we find
\be\label{eq:Svc}\frac{\sv}{\sv_0}&\simeq&S(v+v_c),\\
\label{eq:vc}v_c&=&\frac{V_0\,\m\,\sv_0}{2\pi}.
\ee
Thus we derive and confirm the use of a regularising velocity\footnote{Note that the $T(v)$ term is neglected in Eqs.~(\ref{eq:Svc}-\ref{eq:vc}); we could keep it by adding to $v_c$ a velocity-dependent imaginary part and letting $S(v)\to S(|v+v_c|)$. However, (i) very close to resonance the $T(v)$ term can be omitted, and (ii) studying the Hulth\'en potential, we find that Eqs.~(\ref{eq:Svc}-\ref{eq:vc}) generally provide a good approximation to the full result for perturbative $\sv_0$.}. 
This can be plugged in the usual formula Eq.~(\ref{eq:usual_formula}) using $S(v)\to S\left(v+ v_c\right)$.

The velocity scale $v_c$ is useful for finding the SE saturation point. To assess this scale for phenomenological applications we examine it for the Yukawa potential $V_Y(r)=-\a e^{-mr}/r$ arising from a light mediator with mass $m\ll M$, where $M=2\m$, again, is the WIMP mass. In this example $V_0\approx \a m$, giving 
\be\label{eq:vcnum}
v_c&=&\frac{\a\,m\,M\,\sv_0}{4\pi}\\
&\approx&2\times10^{-3}\,\a\left(\frac{M}{\rm 10~TeV}\right)^2\left(\frac{m/M}{0.1}\right)\left(\frac{\sv_0}{3\times10^{-26}~{\rm cm^3/sec}}\right)\;\;\;\;\;\;\;\;\;\;\;({\rm Yukawa\;potential}).\nonumber\ee
We scaled the high-velocity annihilation cross section to the relevant value for thermal freeze-out dark matter. 

For parameters in the ballpark chosen in Eq.~(\ref{eq:vcnum}) with $\alpha=\mathcal{O}(1)$, the numerical value of $v_c$ makes it phenomenologically relevant for indirect searches for annihilating dark matter, implying that the SE would be saturated in Galactic systems. For comparison, the typical virial velocity in the Milky Way halo is about a factor of 3 smaller than this $v_c$, while in dwarf galaxies the WIMP velocity would be much smaller. Models where these parametrics apply include, for example, the dilaton portal of~\cite{Blum:2014jca}.

For the electroweak supersymmetric WIMPs of, e.g., Refs.~\cite{Hisano:2002fk, Hisano:2003ec, Hisano:2004ds}, we should use $\a\sim10^{-2}$, $M=\mathcal{O}(1~{\rm TeV})$, and $m/M\sim0.1$. This gives $v_c\sim10^{-7}$, well below the relevant velocity scale in low redshift astrophysical systems of interest for indirect detection.

In the early universe, before the dark matter forms virialised halos, the WIMP velocity may be much lower than in a typical Galaxy and the regularization may be important even for TeV-scale or lighter WIMPs with perturbative interactions. However, away from resonances, the saturation velocity of the usual $S(v)$ factor ($v\sim m/\alpha M$) tends to greatly exceed the regularization velocity $v_c$, for $\alpha \ll 1$. Accordingly, for perturbative interactions, pronounced effects of the regularization in the early universe will be confined to the small regions of parameter space near resonance peaks. 

It has been previously noted that in such resonant regions, annihilation may \emph{recouple} during the radiation-dominated epoch after the dark matter kinetically decouples from the Standard Model, since the $\sim 1/v^2$ scaling of $S(v)$ leads to an increasing annihilation rate per comoving volume \cite{Zavala:2009mi}. This effect can lead to a second phase of exponential depletion of the dark matter density, meaning that the correct relic density can be obtained with a much smaller short-range annihilation cross section. In this case, it can be very important to take the regularization into account; examples that we have calculated indicate that the short-range annihilation cross section required to yield the correct relic density in such a resonant scenario can increase by several orders of magnitude, relative to the case with no regularization, albeit in all cases it remains well below the ``standard'' thermal relic value of $\langle \sigma v \rangle \approx 3 \times 10^{-26}$ cm$^3$/s.


\section{Examples}\label{sec:examples}
\subsection{Hulth\'en potential}\label{sssec:hulthen}
The Hulth\'en potential is 
\begin{align}
\label{eq:hulthen}V(r) = -\frac{\a m_* e^{-m_* r}}{1-e^{-m_* r}}.
\end{align}
We consider this potential because, choosing $m_*=(\pi^2/6)m$, it provides a good approximation of the commonly encountered Yukawa potential $V_Y(r)=-\alpha e^{-mr}/r$ with consistent asymptotic behaviour at large and small $r$.

The SE with Eq.~(\ref{eq:hulthen}) admits an analytic solution (see Refs.~\cite{Cassel:2009wt, Slatyer:2009vg} for a computation of the  $S(v)$ term).
Here we add the ingredients needed for the full formula Eq.(\ref{eq:sigma_ann1}). 
Define dimensionless variables,
\begin{align}
\d = \frac{m_*}{2\a\m},
\qquad
\hat p = \frac{p}{2\a\m},
\qquad
\hat r = 2\a\m r,
\qquad
\a_\pm = \frac{i\hat p \pm \sqrt{\d - \hat p^2}}{\d}.
\end{align}
Using these, $g_p(r)$ is written as
\begin{align}
g_p(r) =
\frac{\G(1-\a_+)\G(1-\a_-)}{\G(1-\a_+-\a_-)}
 e^{ipr}~_2F_1(-\a_-,-\a_+,1-\a_+-\a_-,e^{-m_* r}),
\end{align}
with $_2F_1(a,b,c,d)$ the Gauss hypergeometric function. 
This gives
\be
\label{eq:h2} d_p     &=& \frac{ \G(1-\a_+)\G(1-\a_-) }{ \G(1-\a_+-\a_-) },\\
\label{eq:h1}T(v)     &= &    -\frac{\a}{v} \left[ H(\a_+) + H(\a_-) + H(-\a_+) + H(-\a_-) -\left\{p\to p_0\right\}\right], \\
\label{eq:h3}    S(v) &= &\frac{2\pi \a}{v}\frac{\sinh(2\pi\hat p /\d) }{ \cosh(2\pi\hat p /\d) - \cos(2\pi\sqrt{\d-\hat p^2}/\d ) },
\ee
where $H(z)$ is the Harmonic Number.
It is easy to verify that $|d_p|^2=\frac{1}{p}{\rm Im}g'_p(0)=S(v)$.

\begin{figure}[htbp]
\begin{center}
\includegraphics[width=0.475\textwidth]{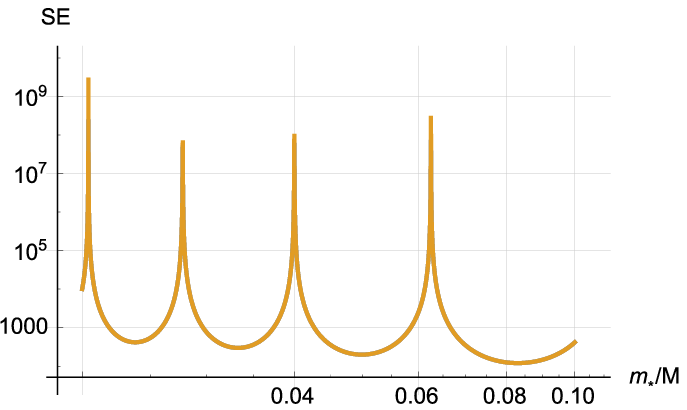}\quad
\includegraphics[width=0.475\textwidth]{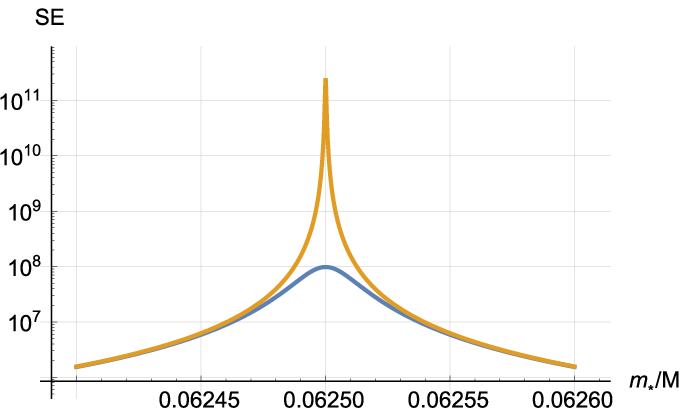}
\caption{SE vs $m_*/M$ for the Hulth\'en potential. {\bf Left}: naive SE given by $S(v)$ of Eq.~(\ref{eq:h3}), corresponding to the usual computation in the literature. The regulated SE computed from Eq.(\ref{eq:sigma_ann1}) is superimposed, but the difference from the usual method is invisible on the scale of the plot. {\bf Right}: zoom-in on one of the resonant SE peaks. The regulated SE obtained from Eq.(\ref{eq:sigma_ann1}) is the smooth lower curve below the sharp peak of the naive $S(v)$. For this plot, we took $\alpha=1$, $v=10^{-6}$, and short-range annihilation and elastic cross sections $\sv_0=1/(32\pi M^2)$ and $\sigma_{sc,0}=(\mu^2\sv_0/4\pi)\sv_0$.}
\label{fig:HY}
\end{center}
\end{figure}

In Fig.~\ref{fig:HY} we show the SE for the Hulth\'en potential. On the left we show the naive SE given by $S(v)$ of Eq.~(\ref{eq:h3}), corresponding to the usual computation in the literature. The regulated SE computed from Eq.(\ref{eq:sigma_ann1}) is superimposed, but the difference from the usual method is invisible on the scale of the plot. On the right we zoom-in on one of the resonant peaks. The regulated SE of Eq.(\ref{eq:sigma_ann1}) is the smooth lower curve. For this plot, we took $\alpha=1$, $v=10^{-6}$, and short-range cross sections $\sv_0=1/(32\pi M^2)$ and ${\rm Re}k_{p_0}=0$ corresponding to short-range elastic cross section $\sigma_{sc,0}=(\mu^2\sv_0/4\pi)\sv_0$.

\begin{figure}[htbp]
\begin{center}
\includegraphics[width=0.6\textwidth]{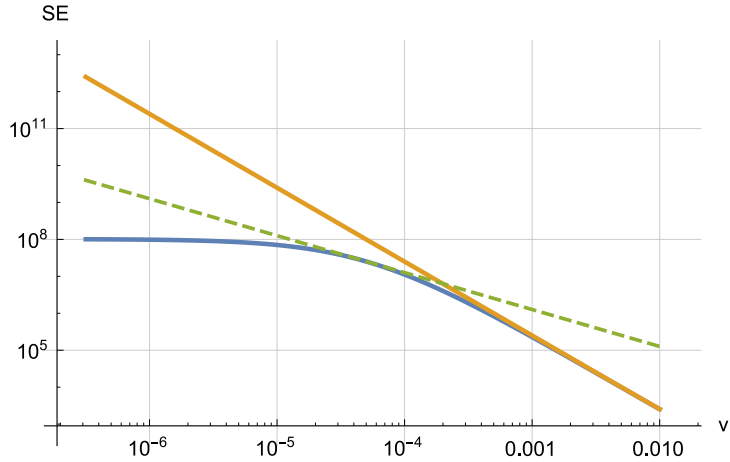}
\caption{SE vs $v$ for the Hulth\'en potential near a resonance peak. The partial-wave unitarity limit for s-wave annihilation, $\sv<\pi/(\mu^2v)$ corresponding to $SE_{max}=\pi/(\mu^2\sv_0v)$, is shown by the dashed line. The naive $S(v)$ goes above the limit at small velocities. The regulated SE saturates the limit and stays below it. For this plot, we took $m_*=0.0625M$, $\alpha=1$, and short-range annihilation and elastic cross sections $\sv_0=1/(32\pi M^2)$ and $\sigma_{sc,0}=(\mu^2\sv_0/4\pi)\sv_0$.}
\label{fig:HYU}
\end{center}
\end{figure}

In Fig.~\ref{fig:HYU} we study the interplay with partial-wave unitarity by inspecting the SE near a resonance peak. The partial-wave unitarity limit for s-wave annihilation, $\sv<\pi/(\mu^2v)$ corresponding to $SE_{max}=\pi/(\mu^2\sv_0v)$, is shown by the dashed line. The naive $S(v)$ goes above the limit at small velocities. The regulated SE saturates the limit and stays below it. For this plot, we took $m_*=0.0625M$, $\alpha=1$, and short-range annihilation and elastic cross sections $\sv_0=1/(32\pi M^2)$ and $\sigma_{sc,0}=(\mu^2\sv_0/4\pi)\sv_0$.

\begin{figure}[htbp]
\begin{center}
\includegraphics[width=0.6\textwidth]{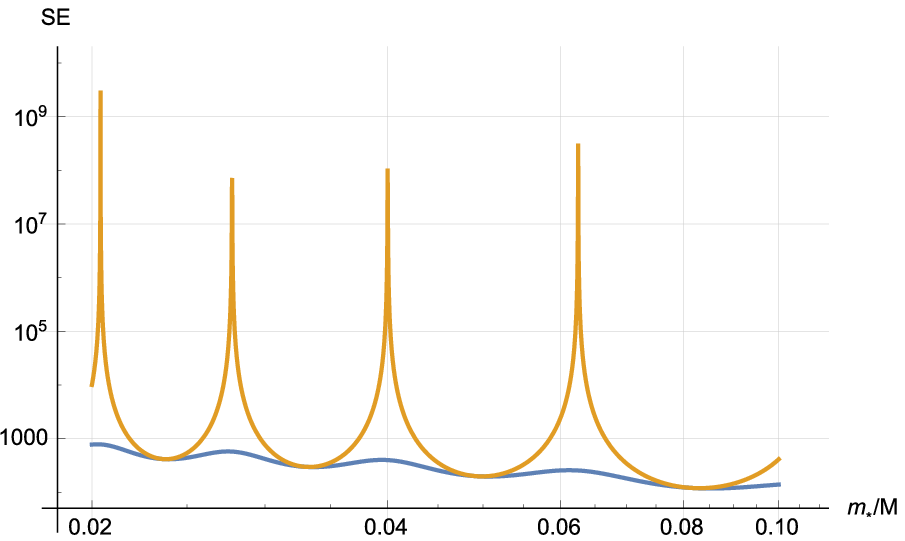}
\caption{SE vs $v$ for the Hulth\'en potential, with a large short-range annihilation cross section. The regulated SE is below the naive sharply-peaked $S(v)$, as the latter violates the unitarity limit over large parametric range even away from a resonance. For this plot, we took $v=10^{-6}$, $\alpha=1$, and short-range annihilation and elastic cross sections $\sv_0=2\pi/M^2$ and $\sigma_{sc,0}=(\mu^2\sv_0/4\pi)\sv_0$.}
\label{fig:HYU2}
\end{center}
\end{figure}

In Fig.~\ref{fig:HYU2} we show SE vs $v$ for the Hulth\'en potential, with a large short-range annihilation cross section. The regulated SE is below the naive sharply-peaked $S(v)$, as the latter violates the unitarity limit over large parametric range even away from a resonance. For this plot, we took $v=10^{-6}$, $\alpha=1$, and short-range annihilation and elastic cross sections $\sv_0=2\pi/M^2$ and $\sigma_{sc,0}=(\mu^2\sv_0/4\pi)\sv_0$.

\subsection{Coulomb potential}\label{sssec:coulomb}
The SE for the attractive Coulomb potential, $V(r)=-\a/r$, can be obtained from the result of Sec.~\ref{sssec:hulthen} by taking the mass parameter $m_*\to0$. 
This leads to
\be
     %
%
\label{eq:c1}T(v)     &= &  \frac{2\a}{v}\left(\log\frac{v_0}{v}-{\rm Re}\psi\left(-\frac{i\a}{v}\right)+{\rm Re}\psi\left(-\frac{i\a}{v_0}\right)\right),\\
\label{eq:c3}    S(v) &= &\frac{2\pi\a}{v\left(1-e^{-\frac{2\pi\a}{v}}\right)},
\ee
where $\psi(z)=\Gamma'(z)/\Gamma(z)$ is the PolyGamma function. We do not quote here the result for the factor $d_p$; the normalization of $d_p$ is given of course by $|d_p|^2=S(v)$, but obtaining the phase requires a modification to the plane wave asymptotic states that we have assumed in Eq.~(\ref{eq:bcd2_for_greenfunction}). This is the standard logarithmic asymptotic phase modification due to the slow-falling $1/r$ potential (see e.g.  \S 7.13 in Ref.~\cite{citeulike:735057}, and \S 135 in Ref.~\cite{Landau:1991wop}). Note that $d_p$ is not needed in the full SE formula for annihilation, Eq.~(\ref{eq:sigma_ann1}), for which one only requires $S(v)$ and $T(v)$.


\subsubsection{Well potential}
Consider the attractive well potential,
\begin{align}
V(r) = -\frac{p_V^2}{2\m}\theta(R-r).
\end{align}
$g_p(r)$ is written as
\begin{align}
g_p(r) = \left\{
\begin{array}{ll}
(1-c_p)e^{i\tilde p r} + c_p e^{-i\tilde p r} & (r < R )\\
d_p e^{ipr} & (r \geq R )
\end{array}
\right.,
\end{align}
where $\tilde p = \sqrt{p^2 + p_V^2}$.
$c_p$ and $d_p$ are determined from continuity of $g_p$ and $g_p'$ at $r=R$.
$g_p'(0)$ is obtained as,
\begin{align}
g_p'(0) = i\tilde p \frac{p\cos \tilde p R - i\tilde p \sin \tilde p R }{ \tilde p\cos \tilde p R - i p \sin \tilde p R }.
\end{align}
Then,
\be
d_p &=& \frac{\tilde p e^{-ipR}}{\tilde p \cos\tilde p R - i p \sin \tilde p R},\\
T(v) &=& \frac{p_V}{p}\left(\frac{p_V}{2\tilde p} \frac{\sin 2\tilde p R}{\cos^2\tilde p R + \frac{p^2}{\tilde p^2}\sin^2 \tilde p R}-\left\{p\to p_0\right\}\right),\\
S(v) &=& \frac{1}{\cos^2\tilde p R + \frac{p^2}{\tilde p^2}\sin^2 \tilde p R }.
\ee
%


%

\section{Conclusion}\label{sec:conc}
Several references in the literature noted that the naive factorized estimate of the Sommerfeld effect (SE), given by $\sv\approx\sv_0\times S(v)$ [see Eq.~(\ref{eq:usual_formula}) and discussion around it], fails in specific parametric points where the Schr\"odinger equation for the long-range force develops a bound state with zero binding energy. At these isolated points, the naive calculation violates partial-wave unitarity. We add to this the observation that a large but finite $S(v)$ factor can boost $\sv$, naively estimated as above, such that it violates the unitarity limit even if the short-range process itself is perturbative, $\sv_0\ll\pi/\m^2v_0$.

Also in the literature, referring specifically to the zero energy bound state problem, it was commented that the apparent unitarity violation should be regularised by the finite width of the zero energy quasi-bound state. In the analyses of Hisano et al~\cite{Hisano:2002fk, Hisano:2003ec, Hisano:2004ds} a full Schr\"odinger equation is derived from the quantum field theory (QFT), containing explicitly the non-hermitian contact term representing short-range annihilation; it is then argued there that the apparent unitarity violation is an artefact of treating the contact term perturbatively in the solution.

We identify the problem from another angle, taking the effective quantum mechanics (QM) perspective where it is clear that the usual formulation of the SE does not conserve probability, because it does not account for the probability flux absorbed at the origin $r\to0$ by the short-range annihilation process. 

We solve the problem by referring to Jackiw's~\cite{Jackiw:1991je} calculation of the scattering amplitude for a delta function potential. In this respect, our contribution is in that we fulfil the suggestion of Hisano et al~\cite{Hisano:2002fk, Hisano:2003ec, Hisano:2004ds} to solve completely the contact term effects. The final expression for the full SE factor for annihilation is given in Eq.~(\ref{eq:sigma_ann1}). 

Calculating the full SE with our formalism is as straightforward as it is in the usual naive formula. A technical  difference is that the full result requires one more quantity encoding the long-range force [contained in the new term $T(v)$, a natural counterpart to the usual $S(v)$ term]; and one more quantity encoding short-range physics (the short-range elastic scattering cross section $\sigma_{sc,0}$). In Sec.~\ref{sec:examples} we gave analytic expressions of the new $T(v)$ term for the Hulth\'en potential (a close approximation to the Yukawa potential) as well as for the Coulomb and well potentials. We also recap the formulae for $S(v)$, found in existing literature for these potentials. 

For perturbative underlying short-range cross section $\sv_0\ll\pi/\mu^2v_0$, under the effect of long-range Yukawa, Hulth\'en, or well potentials, a particularly simple prescription is given in Sec.~\ref{ssec:vc}: the full SE is obtained by the velocity shift $S(v)\to S(v+v_c)$ with $v_c$ given in Eq.~(\ref{eq:vc}).

While we have clarified and solved the basic problem with the SE at large amplification, several issues were left out in the current paper and merit further work. These include (i) generalization to the multi-state problem, relevant for e.g. the electroweak super-partner scenario of~\cite{Hisano:2002fk, Hisano:2003ec, Hisano:2004ds}; (ii) generalization to p-wave and higher-$l$ initial states; (iii) complete characterization of the SE for elastic scattering.

\section*{Acknowledgements}
We thank Michael Dine, Zohar Komargodski, Yosef Nir, and Josef Pradler for comments on the manuscript.
We also thank Martin J. Savage for pointing out relevant nuclear physics work.
The work of R.S. is supported in part by JSPS Research Fellowships for Young Scientists. This work is supported by the U.S. Department of Energy under grant Contract Numbers DE$-$SC00012567 and DE$-$SC0013999.

\appendix

\bibliography{ref}
\bibliographystyle{utphys}

\end{document}